\documentclass[preprint,showpacs,preprintnumbers,amsmath,amssymb]{revtex4-1}

\usepackage{amsmath}
\usepackage{amssymb}
\usepackage{amsthm}
\usepackage{amsfonts}
\usepackage{graphicx}
\usepackage{dcolumn}
\usepackage{bm}
\usepackage{epsfig}
\usepackage{mathrsfs}
\usepackage{tikz}

\begin{document}
\title{Asymptotic correlation functions and FFLO signature  for the one-dimensional attractive spin-1/2 Fermi gas}

\author{J. Y. Lee and X. W. Guan }

 \affiliation{Department of Theoretical Physics,
 Research School of Physics and Engineering,
 Australian National University, Canberra ACT 0200, Australia}

\date{\today}

\begin{abstract}
We investigate the long distance asymptotics of various correlation
functions for the one-dimensional spin-1/2 Fermi gas  with
attractive interactions using the dressed charge formalism. In the
spin polarized phase, these correlation functions exhibit spatial
oscillations with a power-law decay whereby their critical exponents
are found through conformal field theory. We show that spatial
oscillations of the leading terms in the pair correlation function
and the spin correlation function solely depend on $\Delta k_F$ and
$2\Delta k_F$, respectively. Here $\Delta k_F
=\pi(n_{\uparrow}-n_{\downarrow})$ denotes the mismatch between the
Fermi surfaces of spin-up and spin-down fermions. Such spatial
modulations are characteristics of a
Fulde-Ferrell-Larkin-Ovchinnikov (FFLO) state. Our key observation
is that backscattering among the Fermi points of bound pairs and
unpaired fermions results in a one-dimensional analog of the FFLO
state and displays a microscopic origin of the FFLO nature.
Furthermore, we show that the pair correlation function in momentum
space has a peak at the point of mismatch between both Fermi
surfaces $k=\Delta k_F$, which has recently been observed in
numerous numerical studies.
\end{abstract}



\maketitle

\section{Introduction}
Bardeen-Cooper-Schrieffer (BCS) theory was formulated over 50 years
ago as a microscopic theory for superconductivity. One of the
ingredients in BCS theory is pairing between electrons with opposite
momenta and spins, i.e., matching between the Fermi energies of
spin-up and spin-down electrons. In the phase where the system is
partially polarized, Fermi energies of spin-up and spin-down
electrons become unequal. This leads to a non-standard form of
pairing which was predicted independently by Fulde and Ferrell
\cite{Fulde1964}, and Larkin and Ovchinnikov \cite{Larkin1965}.
Fulde and Ferrell discovered that under a strong external field,
superconducting electron pairs have nonzero pairing momentum and
spin polarization. At about the same time, Larkin and Ovchinnikov
suggested that the formation of pairs of electrons with different
momenta, i.e., $\vec{k}$ and $-\vec{k}+\vec{q}$ where $\vec{q}\neq
0$, is energetically favored over pairs of electrons with opposite
momenta, i.e., $\vec{k}$ and $-\vec{k}$, when the separation between
Fermi surfaces is sufficiently large. Consequently, the density of
spins and the superconducting order parameter become periodic
functions of the spatial coordinates. This non-conventional
superconducting state is known in literature as the
Fulde-Ferrell-Larkin-Ovchinnikov (FFLO) state.

More recently, theoretical predictions of the existence of an FFLO
state in one-dimensional (1D) interacting fermions
\cite{Yang1967,Gaudin1967} have emerged by employment of various
methods, such as Bethe ansatz (BA) \cite{Orso,HuHui}, density-matrix
renormalization group (DMRG)
\cite{Feiguin2007,Rizzi2008,Tezuka2008,Meisner,Luscher}, quantum Monte Carlo
(QMC) \cite{Batrouni2008}, mean field theory
\cite{Kinnunen2006,Liu2008,Zhao2008,Cooper}  and bosonization
\cite{Yang2001}. At finite magnetization, it was found  by Feiguin and 
Heidrich-Meisner  \cite{Feiguin2007} that pair
correlations for the attractive Hubbard model in a parabolic
trapping potential has a power-law decay of the form
$n^{\mathrm{pair}}\propto\cos(k_{\mathrm{FFLO}}|x|)/|x|^{\alpha}$
and the momentum pair distribution   has peaks at the mismatch of the Fermi surfaces $k_{\mathrm{FFLO}}=\pi(n_{\uparrow}-n_{\downarrow})$. Wave numbers for the oscillations were
numerically found as $\pi(n_{\uparrow}-n_{\downarrow})$ for the pair
correlation function and as $2\pi(n_{\uparrow}-n_{\downarrow})$ for
the density difference $\langle n_{\uparrow}-n_{\downarrow}\rangle$
\cite{Tezuka2008}. The FFLO pairing wave number was also confirmed
by the occurrence of a peak in the pair momentum distribution
corresponding to the difference between the Fermi momenta of
individual species \cite{Rizzi2008,Batrouni2008}. From mean field
theory, it was demonstrated that the FFLO phase exists in the
large-scale response of the Fermi gas \cite{Cooper} and even for
temperatures up to $0.1T_{F}$ \cite{Liu2008}.

On the other hand, critical behavior of 1D many-body systems with
linear dispersion in the vicinities of their Fermi points can be
described by conformal field theory.  Some time ago, the critical
behavior of the Hubbard model with attractive interaction was
investigated by Bogoliubov and Korepin
\cite{Bogoliubov1988,Bogoliubov1989,Bogolyubov1990,Bogolyubov1992}.
They showed that 1D superconductivity occurs when the average
distance between electron pairs is larger than the average distance
between individual electrons of these pairs. This means that the
correlation function for the single particle Green's function decays
exponentially, i.e.,
$\langle\psi_{n,s}^{\dagger}\psi_{1,s}\rangle\rightarrow e^{-
n/\xi}$ with $\xi=v_F/\Delta$ and $s=\uparrow,\, \downarrow$,
whereas the singlet pair correlation function decays as a power of
distance, i.e.,
$\langle\psi_{n,\uparrow}^{\dagger}\psi_{n,\downarrow}^{\dagger}\psi_{1,\uparrow}\psi_{1,\downarrow}\rangle\rightarrow
n^{-\theta}$. Here $\Delta$ is the energy gap, and the critical
exponents $\xi$ and $\theta$ are both greater than zero. This
criterion is met when the external magnetic field is small, i.e.,
$H<H_{c}$. Once the external field exceeds the critical value, i.e.,
$H>H_{c}$, Cooper pairs are destroyed. Thus both of these
correlation functions decay as a power of distance and the pairs
lose their dominance, i.e., electrons become more or less
independent of each other.

So far, theoretical confirmation of the FFLO state in 1D still
relies on numerical evidence of spatial oscillations in the pair
correlations. Despite key features of the $T=0$ phase diagram
\cite{Orso,HuHui,Guan2007,Mueller,Kakashvili,Wadati} for the
attractive Fermi gas were experimentally confirmed using
finite temperature density profiles of trapped fermionic ${}^6$Li
atoms \cite{Liao2009}, the unambiguous theoretical confirmation and
experimental observation of FFLO pairing is still an open problem.
As remarked in Ref. \cite{Rizzi2008} that the 1D FFLO scenario
proposed in Ref. \cite{Yang2001} does not apply to 1D attractive
fermions where quantum phase transition from the fully-paired phase
into the spin polarized phase does not belong to
commensurate-incommensurate university class, also see Refs.
\cite{Penc,Guan2007}. For 1D attractive spin-1/2 fermions with
polarization \cite{Yang1967,Gaudin1967}, the low-energy physics of
the homogeneous system is described by a two-component
Tomonaga-Luttinger liquid (TLL) of bound pairs and excess unpaired
fermions in the charge sector and ferromagnetic spin-spin
interactions in the spin sector \cite{Erhai}. In this paper, we
determine the critical behavior of the single particle Green's
function, pair correlation function and spin correlation function
within the context of a TLL. We show that the long distance
asymptotics of various correlation functions provide a microscopic
origin of FFLO pairing for 1D attractive fermions.

This paper is organized as follows. We derive finite-size
corrections for the ground state energy of the system in Section
\ref{sec:Finite-size_GS}. In Section \ref{sec:Finite-size_Ex}, we
derive finite-size corrections for low-lying excitations and
introduce the dressed charge formalism. Integral equations for each
component of the dressed charge matrix is solved analytically in the
strong coupling limit $|c|\gg 1$. In Section \ref{sec:Corr-func}, we
derive correlation functions for different operators and discuss the
signature of FFLO pairing. Finally, conclusions and remarks are made
in Section \ref{sec:Conclusion}.

\section{Ground state and finite-size corrections}
\label{sec:Finite-size_GS} We consider $N_{f}$ fermions with $SU(2)$
spin symmetry in a 1D system of length $L$ with periodic boundary
conditions. The Hamiltonian for the spin-1/2 Fermi gas
\cite{Yang1967,Gaudin1967} is given by
\begin{equation}
H=-\sum_{j=1}^{N_{f}}\frac{\partial^{2}}{\partial
x_{j}^{2}}+2c\sum_{1\leq j< k\leq N_{f}}\delta(x_{j}-x_{k}),
\end{equation}
where $c<0$ is the attractive interaction strength. This model is
one of the most important exactly solvable quantum many-body
systems. In recent years, it has attracted considerable attention
from theory \cite{Orso,HuHui,Guan2007,Mueller,Kakashvili,Wadati} and
experiment \cite{Liao2009} due to evidence of the FFLO state.
Systems exhibiting novel phase transitions at $T=0$ are particularly
useful in studying TLL physics \cite{Erhai} and the nature of the
FFLO state.

The quasimomenta for unpaired fermions and bound pairs are given by
$k_{j}$ and $\Lambda_{\alpha}\pm \mathrm{i}c'$ which satisfy the BA
equations
\begin{eqnarray}
k_{j}L &=& 2\pi
I_{j}+\sum_{\alpha=1}^{N_{b}}2\tan^{-1}\left(\frac{k_{j}-\Lambda_{\alpha}}{|c'|}\right), \\
2\Lambda_{\alpha}L &=& 2\pi
J_{\alpha}+\sum_{j=1}^{N_{u}}2\tan^{-1}\left(\frac{\Lambda_{\alpha}-k_{j}}{|c'|}\right)
+\sum_{\beta=1}^{N_{b}}2\tan^{-1}\left(\frac{\Lambda_{\alpha}-\Lambda_{\beta}}{2|c'|}\right),
\end{eqnarray}
where quantum numbers $I_{j}$ and $J_{\alpha}$ are given by
\begin{equation}
I_{j}\equiv\frac{N_{b}}{2} \quad(\mathrm{mod}\phantom{a}1), \qquad
J_{\alpha}\equiv\frac{N_{u}-N_{b}+1}{2}\quad(\mathrm{mod}\phantom{a}1).
\label{eq:IandJ}
\end{equation}
Here $c'=c/2$, and $N_{u}$ and $N_{b}$ denote the number of unpaired
fermions and bound pairs, respectively. The energy and momentum for
this system reads
\begin{equation}
E=\sum_{j=1}^{N_{u}}k_{j}^{2}+\sum_{\alpha=1}^{N_{b}}2(\Lambda_{\alpha}^{2}-|c'|^{2}),
\qquad
P=\sum_{j=1}^{N_{u}}k_{j}+2\sum_{\alpha=1}^{N_{b}}\Lambda_{\alpha}.
\end{equation}

We define monotonic increasing counting functions
$z_{u}^{L}(k_{j}):=I_{j}/L$ and
$z_{b}^{L}(\Lambda_{\alpha}):=J_{\alpha}/L$ and re-label the
variables $k\rightarrow k_{u}$, $\lambda\rightarrow k_{b}$,
$I_{j}\rightarrow I_{u,j}$ and $J_{\alpha}\rightarrow I_{b,\alpha}$
so that we can express the root densities in a general form as
\begin{eqnarray}
\rho_{u}^{L}(k_{u}) &:=& \frac{d}{dk_{u}}z_{u}^{L}(k_{u})
=\frac{1}{2\pi}-\frac{1}{L}\sum_{\alpha=1}^{N_{b}}a_{1}(k_{u}-k_{b,\alpha}),\label{eq:rho_u}
\\ \rho_{b}^{L}(k_{b}) &:=& \frac{d}{dk_{b}}z_{b}^{L}(k_{b})
=\frac{1}{\pi}-\frac{1}{L}\sum_{j=1}^{N_{u}}a_{1}(k_{b}-k_{u,j})-\frac{1}{L}\sum_{\beta=1}^{N_{b}}a_{2}(k_{b}-k_{b,\beta}),\label{eq:rho_b}
\end{eqnarray}
where $a_{n}(k)$ is defined by
\begin{equation}
a_{n}(k)=\frac{1}{\pi}\frac{n|c'|}{(nc')^{2}+k^{2}}.
\end{equation}
Here $k_{\alpha,j}$ (for $j=1,2,\ldots,N_{\alpha}$ and $\alpha=u,b$)
denote the BA roots for unpaired fermions and bound pairs in the
ground state.

Using the Euler-Maclaurin formula for contributions up to
$O(1/L^{2})$ when $L\gg 1$,  the finite-size corrections to the root
densities can be written in the generic form as
\begin{eqnarray}
\nonumber \rho_{\alpha}^{L}(k_{\alpha}) &=&
\rho_{\alpha}^{(0)}(k_{\alpha})+\sum_{\beta=u,b}\int_{-Q_{\beta}}^{Q_{\beta}}
K_{\alpha\beta}(k_{\alpha}-k_{\beta})\rho_{\beta}^{L}(k_{\beta})dk_{\beta}
\\ && +\frac{1}{24L^{2}}\sum_{\beta=u,b}\left[\frac{K'_{\alpha\beta}(k_{\alpha}-Q_{\beta})}{\rho_{\beta}^{L}(Q_{\beta})}
-\frac{K'_{\alpha\beta}(k_{\alpha}+Q_{\beta})}{\rho_{\beta}^{L}(-Q_{\beta})}\right],
\qquad (\alpha=u,b) \label{eq:rho_gen}
\end{eqnarray}
where
\begin{equation}
\left(
  \begin{array}{c}
    \rho_{u}^{(0)}(k_{u}) \\
    \rho_{b}^{(0)}(k_{b}) \\
  \end{array}
\right)=\left(
          \begin{array}{c}
            1/2\pi \\
            1/\pi \\
          \end{array}
        \right), \qquad \mathbf{K}(k)=\left(
                          \begin{array}{cc}
                            K_{uu}(k) & K_{ub}(k) \\
                            K_{bu}(k) & K_{bb}(k) \\
                          \end{array}
                        \right)=\left(
                                  \begin{array}{cc}
                                    0 & -a_{1}(k) \\
                                    -a_{1}(k) & -a_{2}(k) \\
                                  \end{array}
                                \right).\label{eq:K_matrix}
\end{equation}
Here, the Fermi points are denoted by $\pm Q_{\alpha}$. Notice that
$\mathbf{K}(k)$ is a symmetric matrix.

In order to calculate finite-size corrections for the ground state
and low energy excitations, we introduce the thermodynamic Bethe
ansatz (TBA) \cite{Y-Y,Takahashi}, which provides a powerful and
elegant way to study the thermodynamics of 1D integrable systems. It
becomes convenient to analyze phase transitions and low-lying
excitations in the presence of external fields at zero temperature.
In the thermodynamic limit, the grand partition function is
$Z=tr(\mathrm{e}^{-\cal{H}/T})=\mathrm{e}^{-G/T}$, where the Gibbs
free energy is given by $G = E - HM^z - \mu n - TS$, and is written
in terms of the magnetization $H$, the chemical potential $\mu$ and
the entropy $S$ \cite{Takahashi}. Equilibrium states satisfy the
condition of minimizing the Gibbs free energy with respect to
particle and hole densities for the charge and spin degrees of
freedom (more details are given in
Refs.~\cite{Lai1971,Lai1973,Takahashi,Schlottmann1993,Guan2007}). At
zero temperature, the ground state properties are determined by the
dressed energy equations
\begin{equation}
\varepsilon_{\alpha}(k_{\alpha})=\varepsilon_{\alpha}^{(0)}(k_{\alpha})+\sum_{\beta=u,b}\int_{-Q_{\beta}}^{Q_{\beta}}
K_{\alpha\beta}(k_{\alpha}-k_{\beta})\varepsilon_{\beta}(k_{\beta})dk_{\beta},
\qquad (\alpha=u,b), \label{eq:dressedEnergy}
\end{equation}
where $\varepsilon_{\alpha}^{(0)}(k_{\alpha})$ are given by
\begin{equation}
\left(
  \begin{array}{c}
    \varepsilon_{u}^{(0)}(k_{u}) \\
    \varepsilon_{b}^{(0)}(k_{b}) \\
  \end{array}
\right)=\left(
          \begin{array}{c}
            k_{u}^{2} \\
            2k_{b}^{2}-|c|^{2}/2 \\
          \end{array}
        \right).
\end{equation}

1D many-body systems are critical at $T=0$ and exhibit not only
global scale invariance but local scale invariance too, i.e.,
conformal invariance. The conformal group is infinite dimensional
and completely determines the conformal dimensions and correlation
functions when the excitations are gapless \cite{Belavin1984}.
Conformal invariance predicts that the energy per unit length has a
universal finite-size scaling form that is characterized by the
dimensionless number $C$, which is the central charge of the
underlying Virasoro algebra \cite{Blote1986,Affleck1986}.  From the
density distributions (\ref{eq:rho_gen}) and dressed energy
equations (\ref{eq:dressedEnergy}), the finite-size corrections to
the ground state energy is given by
\begin{equation}
\varepsilon_{0}=\varepsilon_{0}^{\infty}-\frac{C\pi}{6L^{2}}\sum_{\alpha=u,b}v_{\alpha},
\label{eq:energy_Lfinalground}
\end{equation}
where $C=1$, and $v_{u}$ and $v_{b}$ are the velocities of unpaired
fermions and bound pairs, respectively. They are defined as
\begin{equation}
v_{\alpha}:=\pm\left.\frac{d\varepsilon_{\alpha}(k_{\alpha})}{dp_{\alpha}(k_{\alpha})}\right|_{k_{\alpha}=\pm
Q_{\alpha}}=\pm\frac{\varepsilon'_{\alpha}(\pm
Q_{\alpha})}{p'_{\alpha}(Q_{\alpha})}=\pm\frac{\varepsilon'_{\alpha}(\pm
Q_{\alpha})}{2\pi\rho_{\alpha}(\pm Q_{\alpha})}, \qquad
(\alpha=u,b),
\end{equation}
where prime denotes the derivative with respect to $k_{\alpha}$ and
$p_{\alpha}(k_{\alpha})=\lim_{L\rightarrow\infty}2\pi
z_{\alpha}^{L}(k_{\alpha})$. The term $\varepsilon_{0}^{\infty}$
represents the ground state energy in the thermodynamic limit, i.e.,
$N,L\rightarrow\infty$. In the strong coupling limit, exact
expressions for the velocities can be found in
Refs.~\cite{Guan2007,Guan2010}.

\section{Low-lying excitations and dressed charge equations}
\label{sec:Finite-size_Ex} Critical phenomena of critical systems
are described by finite-size corrections for their low-lying
excitations. The method we use to study correlation functions of the
spin-1/2 Fermi gas with attractive interaction follows
closely the method set out in
Refs.~\cite{Woynarovich1989,Kawakami1991,Frahm1991,Hubbardbook}. The
conformal dimensions of two-point correlation functions can be
calculated from the elements of the dressed charge matrix
$\mathbf{Z}$. Long distance asymptotics of various correlation
functions are then examined through the dressed charge formalism at
the $T=0$. Three types of low-lying excitations are considered in
the calculations of finite-size corrections.

Type 1 excitation is characterized by moving a particle close to the
right or left Fermi points outside the Fermi sea. It is equivalent
to changing the quantum numbers $I_{\alpha,j}$ close to
$I_{\alpha}^{\pm}$ for unpaired fermions ($\alpha=u$) and bound
pairs ($\alpha=b$). $I_{\alpha}^{\pm}$ characterize the Fermi points
of each Fermi sea and are given by
$I_{\alpha}^{+}=I_{\alpha}^{\mathrm{max}}+1/2$ and
$I_{\alpha}^{-}=I_{\alpha}^{\mathrm{min}}-1/2$. The change in total
momentum from Type 1 excitations is
\begin{equation}
\Delta
P=\frac{2\pi}{L}\sum_{\alpha=u,b}(N_{\alpha}^{+}-N_{\alpha}^{-}),
\end{equation}
and the change in energy is
\begin{eqnarray}
\nonumber \Delta E &=&
\frac{2\pi}{L}\sum_{\alpha=u,b}\frac{\varepsilon'_{\alpha}(Q_{\alpha}|Q^{\pm})}
{p'_{\alpha}(Q_{\alpha}|Q^{\pm})}(N_{\alpha}^{+}+N_{\alpha}^{-}) \\
&=&
\frac{2\pi}{L}\sum_{\alpha=u,b}v_{\alpha}(N_{\alpha}^{+}+N_{\alpha}^{-}).
\end{eqnarray}
Here $N_{\alpha}^{+}\geq 0$ ($N_{\alpha}^{-}\geq 0$) stems from the
change in distribution of quantum numbers close to the right (left)
Fermi points. This type of excitation is commonly known as
particle-hole excitation.

Type 2 excitation arises from the change in total number of unpaired
fermions or bound pairs. It is characterized by the change in
quantum numbers
\begin{equation}
N_{\alpha}=I_{\alpha}^{+}-I_{\alpha}^{-}, \qquad (\alpha=u,b),
\end{equation}
i.e., $\Delta
N_{\alpha}=N_{\alpha}^{\mathrm{excited}}-N_{\alpha}^{\mathrm{ground}}$.

On the other hand, Type 3 excitation is caused by moving a particle
from the left Fermi point to the right Fermi point and vice versa.
This type of excitation is also known as backscattering. It is
characterized by the quantum numbers
\begin{equation}
\Delta D_{\alpha}=\frac{I_{\alpha}^{+}+I_{\alpha}^{-}}{2}, \qquad
(\alpha=u,b), \label{eq:Dalpha}
\end{equation}
while leaving $\Delta N_{\alpha}$ unchanged.

All three types of excitations can be unified in the following form
of the finite-size corrections for the energy and total momentum of
the system
\begin{eqnarray}
\Delta E &=& \frac{2\pi}{L}\left(\frac{1}{4}\phantom{.}^{t}(\Delta
N)^{t}(\mathbf{Z}^{-1})\mathbf{VZ}^{-1}\Delta N+
\phantom{.}^{t}(\Delta D)\mathbf{ZV}^{t}\mathbf{Z}\Delta
D+\sum_{\alpha=u,b}v_{\alpha}(N_{\alpha}^{+}+N_{\alpha}^{-})\right),
\label{eq:energy_Lfinal}
\\ \Delta P &=& \frac{2\pi}{L}\left(\phantom{.}^{t}\Delta N\Delta D+N_{u}\Delta
D_{u}+N_{b}\Delta
D_{b}+\sum_{\alpha=u,b}v_{\alpha}(N_{\alpha}^{+}-N_{\alpha}^{-})\right).
\end{eqnarray}
Here we use the notations
\begin{eqnarray}
&& \nonumber \Delta N=\left(
           \begin{array}{c}
             \Delta N_{u} \\
             \Delta N_{b} \\
           \end{array}
         \right),\qquad \Delta D=\left(
                                   \begin{array}{c}
                                     \Delta D_{u} \\
                                     \Delta D_{b} \\
                                   \end{array}
                                 \right),\\ &&\mathbf{V}=\left(
                                                             \begin{array}{cc}
                                                               v_{u} & 0 \\
                                                               0 & v_{b} \\
                                                             \end{array}
                                                           \right),\qquad
                                                           \mathbf{Z}=\left(
                                                                        \begin{array}{cc}
                                                                          Z_{uu}(Q_{u}) & Z_{ub}(Q_{b}) \\
                                                                          Z_{bu}(Q_{u}) & Z_{bb}(Q_{b}) \\
                                                                        \end{array}
                                                                      \right).
\end{eqnarray}
The dressed charge equations are a set of four coupled integral
equations that read
\begin{eqnarray}
Z_{uu}(k) &=&
1-\int_{-Q_{b}}^{Q_{b}}a_{1}(k-\lambda)Z_{ub}(\lambda)d\lambda, \label{eq:dressed_uu} \\
Z_{ub}(k) &=&
-\int_{-Q_{u}}^{Q_{u}}a_{1}(k-\lambda)Z_{uu}(\lambda)d\lambda-\int_{-Q_{b}}^{Q_{b}}a_{2}(k-\lambda)Z_{ub}(\lambda)d\lambda,
\label{eq:dressed_ub}
\\ Z_{bu}(k) &=&
-\int_{-Q_{b}}^{Q_{b}}a_{1}(k-\lambda)Z_{bb}(\lambda)d\lambda, \label{eq:dressed_bu} \\
Z_{bb}(k) &=&
1-\int_{-Q_{u}}^{Q_{u}}a_{1}(k-\lambda)Z_{bu}(\lambda)d\lambda-\int_{-Q_{b}}^{Q_{b}}a_{2}(k-\lambda)Z_{bb}(\lambda)d\lambda.
\label{eq:dressed_bb}
\end{eqnarray}
Quantum numbers $\Delta D_{u}$ and $\Delta D_{b}$ (\ref{eq:Dalpha})
are chosen based on the conditions given in Eq.~\eqref{eq:IandJ} and
also on the conditions that $\Delta D_{u}\equiv\Delta N_{u}/2$ (mod
1) and $\Delta D_{b}\equiv\Delta N_{b}/2$ (mod 1). Combining both
conditions together with the definition given in
Eq.~\eqref{eq:Dalpha} yields
\begin{equation}
\Delta D_{u}\equiv\frac{\Delta N_{u}+\Delta
N_{b}}{2}\quad\mathrm{(mod\phantom{a}1)},\qquad \Delta
D_{b}\equiv\frac{\Delta N_{u}}{2}\quad\mathrm{(mod\phantom{a}1)}.
\end{equation}

When the external magnetic field $H$ is smaller than the critical
field, spin excitations for this model are gapped. Once $H$ exceeds
this critical field, spin excitations become gapless and the system
becomes conformally invariant. In this spin polarized phase, spin
degrees of freedom are suppressed due to the ferromagnetic nature of
excess unpaired fermions under a magnetic field. Therefore, bound
pairs and excess unpaired fermions form two Fermi seas which can be
described by a two-component TLL at low temperatures. Hence
conformal invariance results in a universal finite-size scaling form
of the energy shown in Eqs.~\eqref{eq:energy_Lfinalground} and
\eqref{eq:energy_Lfinal}, and a universal form of the critical
exponents of two-point correlation functions between primary fields
$\langle O^{\dagger}(x,t)O(x',t')\rangle$ which are determined by
the finite-size corrections of the model. Multi-point correlation
functions can be derived by taking the product of two-point
correlation functions.

When $T=0$, the correlation functions of 1D systems decay as the
power of distance, but when $T>0$ they decay exponentially.
Following the standard calculations in Ref.~\cite{Hubbardbook}, the
conformal dimensions are given by
\begin{eqnarray}
2\Delta_{u}^{\pm} &=& \left(Z_{uu}\Delta D_{u}+Z_{bu}\Delta D_{b}\pm
\frac{Z_{bb}\Delta
N_{u}-Z_{ub}\Delta N_{b}}{2\det Z}\right)^{2}+2N_{u}^{\pm}, \\
2\Delta_{b}^{\pm} &=& \left(Z_{ub}\Delta D_{u}+Z_{bb}\Delta D_{b}\pm
\frac{Z_{uu}\Delta N_{b}-Z_{bu}\Delta N_{u}}{2\det
Z}\right)^{2}+2N_{b}^{\pm},
\end{eqnarray}
where $N_{\alpha}^{\pm}$ ($\alpha=u,b$) characterize the descendent
fields from the primary fields. General two-point correlation
functions at $T=0$ take the form
\begin{equation}
\langle O(x,t)O(0,0)\rangle=\frac{\exp(-2\mathrm{i}(N_{u}\Delta
D_{u}+N_{b}\Delta D_{b})x)}
{(x-\mathrm{i}v_{u}t)^{2\Delta^{+}_{u}}(x+\mathrm{i}v_{u}t)^{2\Delta^{-}_{u}}
(x-\mathrm{i}v_{b}t)^{2\Delta^{+}_{b}}(x+\mathrm{i}v_{b}t)^{2\Delta^{-}_{b}}}.
\label{eq:corrfunc_gen}
\end{equation}
The exponential oscillating term in the asymptotic behavior comes
from Type 3 excitations, i.e., backscattering. Quantum numbers for
the low-lying excitations completely determine the nature of the
asymptotic behavior of these correlations. Here we are only
concerned with the $T=0$ case.

The four dressed charge equations can be broken up into sets of two
pairs. Eqs.~\eqref{eq:dressed_uu} and \eqref{eq:dressed_ub}
constitute one pair, whilst Eqs.~\eqref{eq:dressed_bu} and
\eqref{eq:dressed_bb} make up the other. Since we are interested in
the strong coupling limit $|c|\gg 1$, both sets of equations can be
solved iteratively up to accuracy $1/|c|$. Let us consider the first
set. Substituting Eq.~\eqref{eq:dressed_uu} into
Eq.~\eqref{eq:dressed_ub} and iterating the terms give
\begin{eqnarray}
\nonumber Z_{ub}(k) &=& -\int_{-Q_{u}}^{Q_{u}}d\lambda
a_{1}(k-\lambda)+\int_{-Q_{b}}^{Q_{b}}d\lambda\int_{-Q_{u}}^{Q_{u}}d\lambda'
a_{2}(k-\lambda)a_{2}(\lambda-\lambda') \nonumber\\
&&  -\int_{-Q_{u}}^{Q_{u}}d\lambda\int_{-Q_{b}}^{Q_{b}}d\lambda'\int_{-Q_{u}}^{Q_{u}}d\lambda^{\prime\prime}
a_{1}(k-\lambda)a_{1}(\lambda-\lambda')a_{1}(\lambda'-\lambda^{\prime\prime})+\ldots
\label{eq:dressed_ub_iter}
\end{eqnarray}
The functions $a_{n}(k)$ have leading order $1/|c|$, hence we can
ignore all terms that have two or more multiples of $a_{n}(k)$. This
procedure yields
\begin{eqnarray}
\nonumber Z_{ub}(Q_{b}) &\approx& -\int_{-Q_{u}}^{Q_{u}}d\lambda
a_{1}(Q_{b}-\lambda) \approx -\frac{4Q_{u}}{\pi |c|}.
\end{eqnarray}
Substituting Eq.~\eqref{eq:dressed_ub_iter} into
Eq.~\eqref{eq:dressed_uu}, we obtain
\begin{eqnarray}
Z_{uu}(Q_{u}) &=&
1+\int_{-Q_{b}}^{Q_{b}}d\lambda\int_{-Q_{u}}^{Q_{u}}d\lambda'
a_{1}(Q_{u}-\lambda)a_{1}(\lambda-\lambda')+\ldots \\ &\approx& 1
\end{eqnarray}

\begin{figure}
\centering
\begin{tabular}{cc}
\epsfig{file=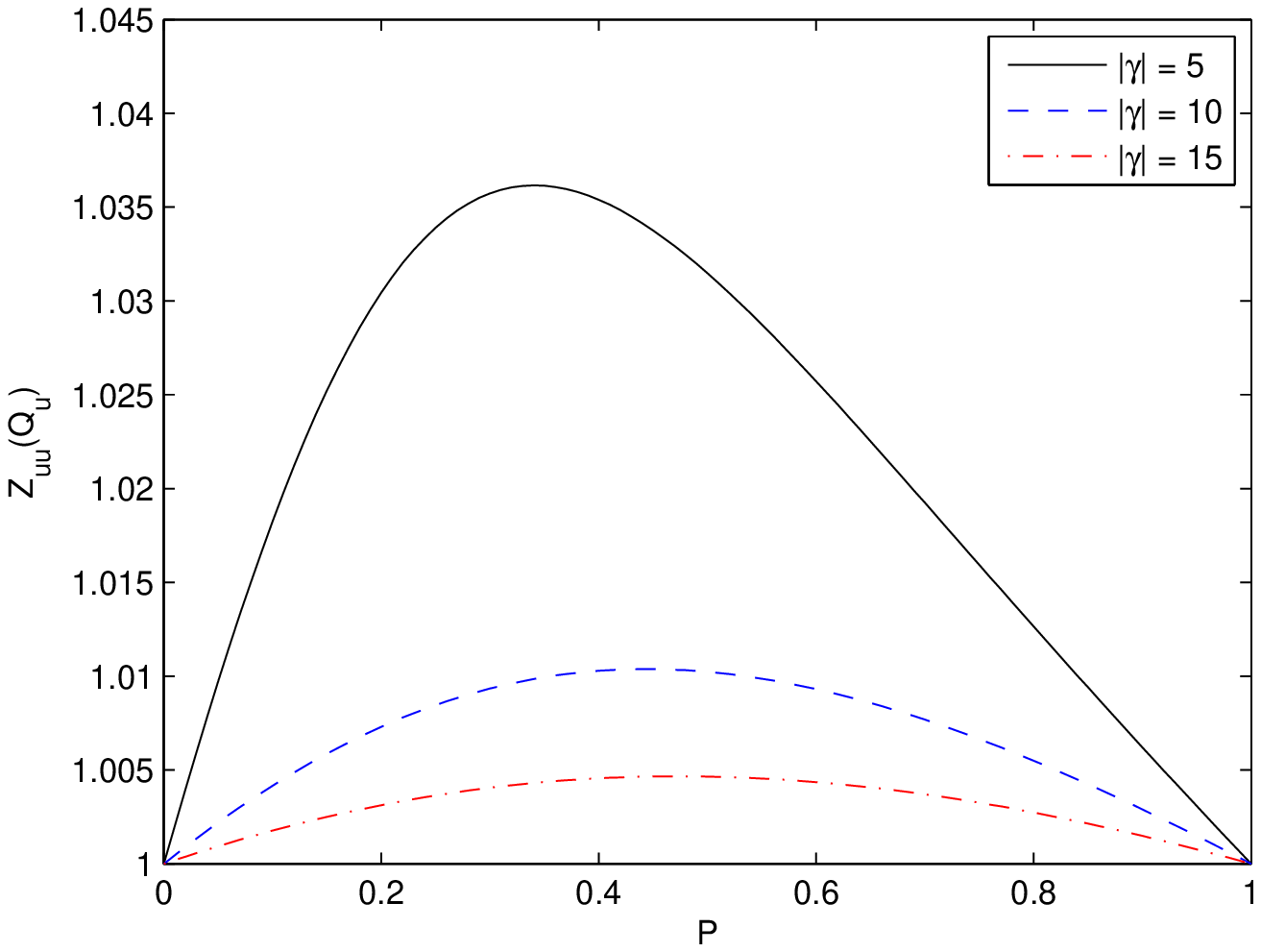,width=0.5\linewidth,clip=} &
\epsfig{file=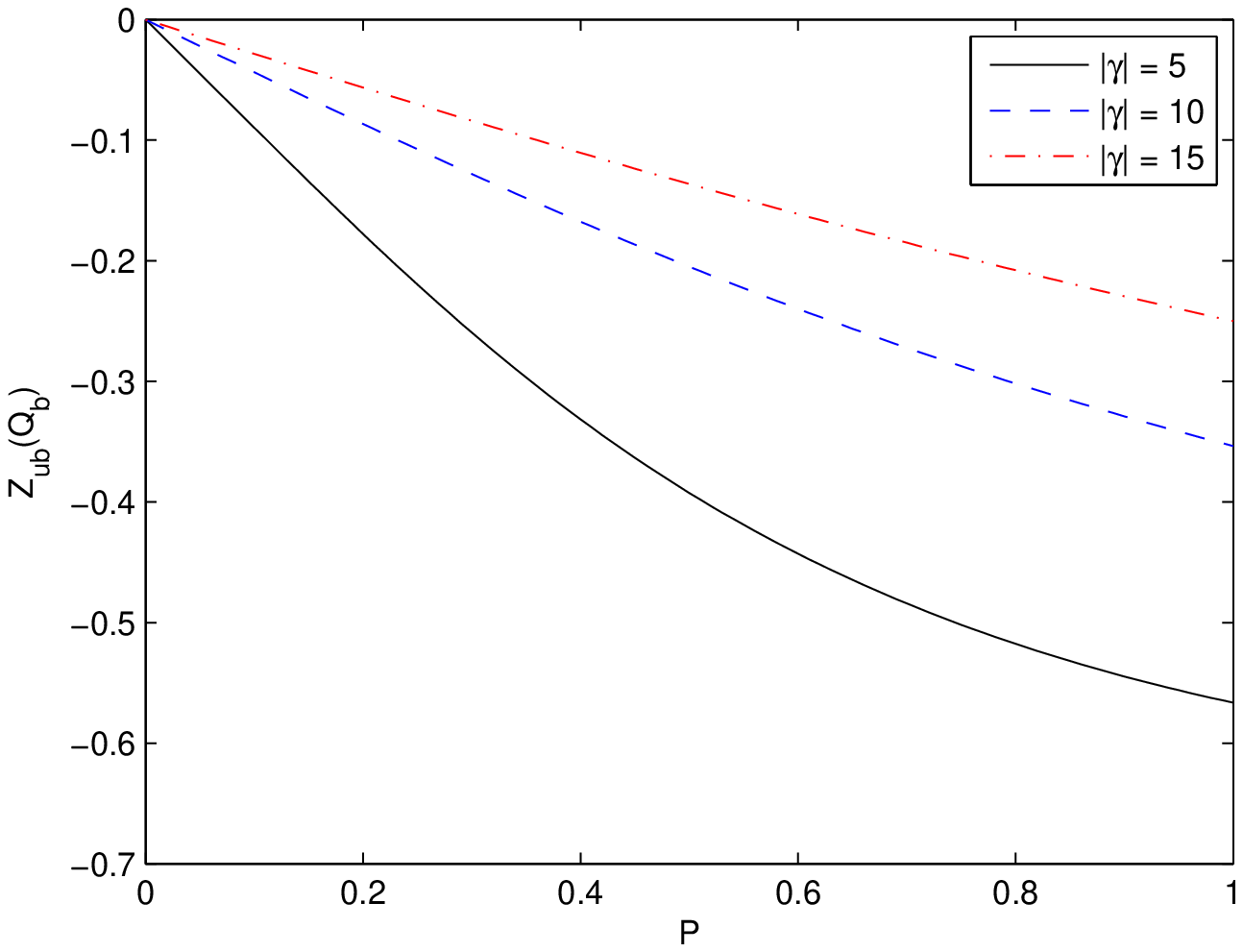,width=0.5\linewidth,clip=} \\
\epsfig{file=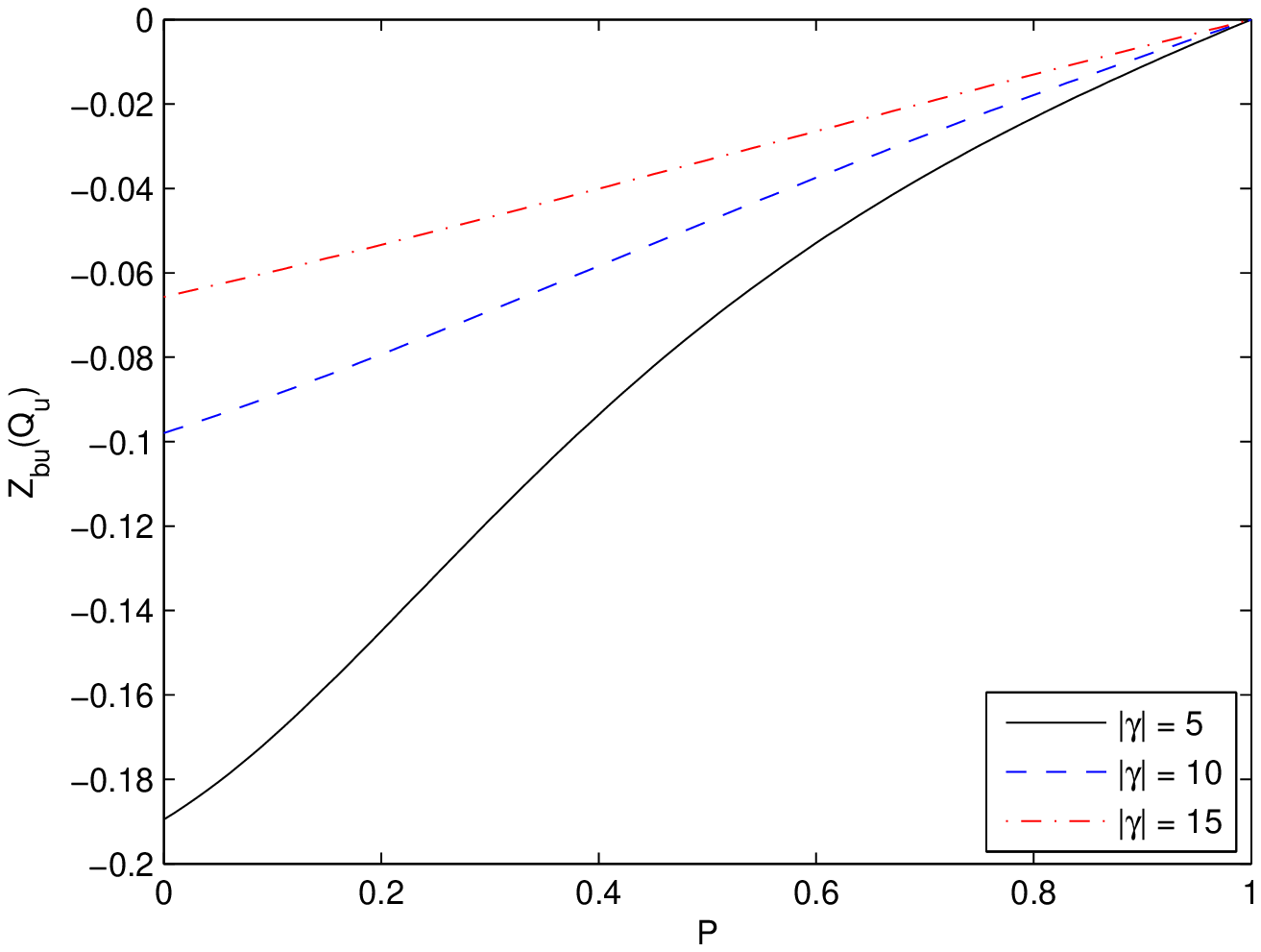,width=0.5\linewidth,clip=} &
\epsfig{file=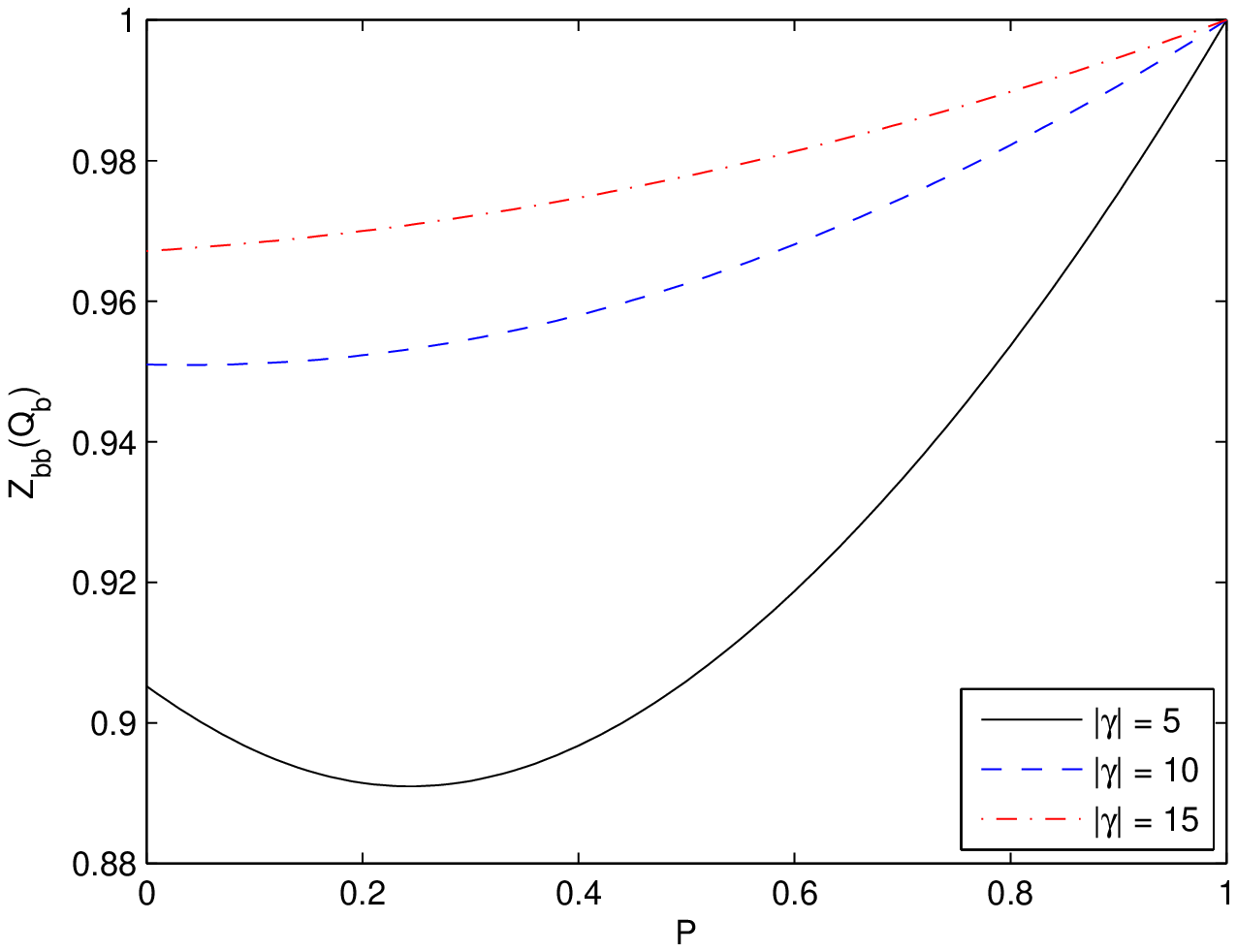,width=0.5\linewidth,clip=}
\end{tabular}\caption{These figures show a plot of the
dressed charges $Z_{uu}(Q_{u})$, $Z_{ub}(Q_{b})$, $Z_{bu}(Q_{u})$
and $Z_{bb}(Q_{b})$ versus polarization for different values of
$|\gamma|$.}\label{fig:Z}
\end{figure}

Next, we consider the second set of equations. Repeating the same
arguments as before, Eq.~\eqref{eq:dressed_bb} at the Fermi point
$Q_{b}$ becomes
\begin{eqnarray}
\nonumber Z_{bb}(Q_{b}) &=& 1-\int_{-Q_{b}}^{Q_{b}}d\lambda
a_{2}(Q_{b}-\lambda)+\int_{-Q_{u}}^{Q_{u}}d\lambda\int_{-Q_{b}}^{Q_{b}}d\lambda'
a_{1}(Q_{b}-\lambda)a_{1}(\lambda-\lambda')
\\ && \nonumber +\int_{-Q_{b}}^{Q_{b}}d\lambda\int_{-Q_{b}}^{Q_{b}}d\lambda'
a_{2}(Q_{b}-\lambda)a_{2}(\lambda-\lambda')+\ldots \\ &\approx&
1-\frac{2Q_{b}}{\pi |c|}.
\end{eqnarray}
Eq.~\eqref{eq:dressed_bu} at the Fermi point $Q_{u}$ then reads
\begin{eqnarray}
\nonumber Z_{bu}(Q_{u}) &=& -\int_{-Q_{b}}^{Q_{b}}d\lambda
a_{1}(Q_{u}-\lambda)+\int_{-Q_{b}}^{Q_{b}}d\lambda\int_{-Q_{b}}^{Q_{b}}d\lambda'
a_{1}(Q_{u}-\lambda)a_{2}(\lambda-\lambda')+\ldots \\ &\approx&
-\frac{4Q_{b}}{\pi |c|}.
\end{eqnarray}

From Ref.~\cite{Guan2007}, the Fermi points in the strongly
attractive limit are given by
\begin{eqnarray}
Q_{u} &\approx& \pi n_{f}P\left(1+\frac{2(1-P)}{|\gamma|}\right), \\
Q_{b} &\approx& \frac{\pi
n_{f}(1-P)}{4}\left(1+\frac{(1-P)}{2|\gamma|}+\frac{2P}{|\gamma|}\right),
\end{eqnarray}
where $n_{f}=N_{f}/L$ is the density of fermions per unit length,
$\gamma=c/n_{f}$ is the dimensionless interaction parameter and
$P=(N_{\uparrow}-N_{\downarrow})/N_{f}=N_{u}/N_{f}$ is the
polarization. Inserting these relations into the expressions for
dressed charges, we obtain
\begin{eqnarray}
Z_{uu}(Q_{u}) &\approx& 1, \qquad Z_{ub}(Q_{b}) \approx
-\frac{4P}{|\gamma|}, \nonumber \\
Z_{bu}(Q_{u}) &\approx& -\frac{(1-P)}{|\gamma|}, \qquad
Z_{bb}(Q_{b}) \approx
1-\frac{(1-P)}{2|\gamma|}.\label{eq:dressedCharges}
\end{eqnarray}
In FIG.~\ref{fig:Z}, the dressed charges are numerically calculated
and plotted against polarization for different values of interaction
strength $|\gamma|$.

In the strong coupling limit, the external magnetic field $H$ is
related to the polarization as
\begin{equation}
H\approx\frac{n^{2}|\gamma|^{2}}{2}+2\pi^{2}n^{2}P^{2}\left(1+\frac{4(1-P)}{|\gamma|}-\frac{4P}{3|\gamma|}\right)
-\frac{\pi^{2}n^{2}(1-P)^{2}}{8}\left(1+\frac{4P}{|\gamma|}\right).
\end{equation}
With this relation, we can evaluate the dressed charges for
different values of $H$. From the expressions for the dressed
charges in Eq.~\eqref{eq:dressedCharges}, the conformal dimensions
$\Delta_{\alpha}^{\pm}$ in terms of polarization are given by
\begin{eqnarray}
2\Delta_{u}^{\pm} &\approx& \left(\Delta D_{u}\pm\frac{\Delta
N_{u}}{2}\right)^{2}-\frac{8P}{|\gamma|}\left(\Delta
D_{u}\pm\frac{\Delta N_{u}}{2}\right)\left(\Delta
D_{b}\mp\frac{\Delta N_{b}}{2}\right)+2N_{u}^{\pm}, \\
\nonumber 2\Delta_{b}^{\pm} &\approx&
\left(1-\frac{(1-P)}{|\gamma|}\right)\left(\Delta
D_{b}\pm\frac{\Delta N_{b}}{2}\right)^{2}
\\ && -\left(\frac{8P}{|\gamma|}\Delta
D_{u}\mp\frac{(1-P)}{|\gamma|}\Delta N_{u}\right)\left(\Delta
D_{b}\pm\frac{\Delta N_{b}}{2}\right)+2N_{b}^{\pm}.
\end{eqnarray}

\section{Correlation functions at zero temperature}
\label{sec:Corr-func} Here we consider 4 types of correlation
functions, namely the single particle Green's function
$G_{\uparrow}(x,t)$, charge density correlation function
$G_{nn}(x,t)$, spin correlation function $G^{z}(x,t)$, and pair
correlation function $G_{p}(x,t)$. Each correlation function is
derived based on the choice of $\Delta N_{u}$ and $\Delta N_{b}$.

The one particle Green's function, which is also called the
Fermi-field (FF) correlation function in some literature, decays
exponentially when the external magnetic field is not strong enough
to overcome the gap associated with the breaking of bound states
\cite{Bogoliubov1988,Bogoliubov1989,Bogolyubov1990,Bogolyubov1992}.
Once in the gapless phase, i.e., when $H_{c1}<H<H_{c2}$ where
$H_{c1}$ and $H_{c2}$ are the critical fields mentioned in
Ref.~\cite{Guan2007}, every correlation function at zero temperature
decays spatially as some form of power law
\cite{Belavin1984,Blote1986,Affleck1986,Cardy1986,Izergin1989}.
$G_{\uparrow}(x,t)$ is characterized by $(\Delta N_{u},\Delta
N_{b})=(1,0)$ which in turn allows quantum numbers $\Delta
D_{u}\in\mathbb{Z}+1/2$ and $\Delta D_{b}\in\mathbb{Z}+1/2$. The
leading terms are then given by
\begin{eqnarray}
\nonumber G_{\uparrow}(x,t) &=&
\langle\psi_{\uparrow}^{\dagger}(x,t)\psi_{\uparrow}(0,0)\rangle
\\ &\approx& \frac{A_{\uparrow,1}\cos\left(\pi(n_{\uparrow}-2n_{\downarrow})x\right)}
{|x+\mathrm{i}v_{u}t|^{\theta_{1}}|x+\mathrm{i}v_{b}t|^{\theta_{2}}}
+\frac{A_{\uparrow,2}\cos\left(\pi
n_{\downarrow}x\right)}{|x+\mathrm{i}v_{u}t|^{\theta_{3}}|x+\mathrm{i}v_{b}t|^{\theta_{4}}},
\end{eqnarray}
where the critical exponents are given by
\begin{eqnarray}
\theta_{1} &\approx& 1+\frac{4P}{|\gamma|}, \qquad  \theta_{2} \approx
\frac{1}{2}-\frac{(1-P)}{2|\gamma|}+\frac{4P}{|\gamma|}, \nonumber \\
\theta_{3} &\approx& 1-\frac{4P}{|\gamma|}, \qquad  \theta_{4} \approx
\frac{1}{2}-\frac{(1-P)}{2|\gamma|}-\frac{4P}{|\gamma|}.
\end{eqnarray}
The first term in $G_{\uparrow}(x,t)$ comes from $(\Delta
D_{u},\Delta D_{b})=(1/2,-1/2)$ and the second term comes from
$(\Delta D_{u},\Delta D_{b})=(1/2,1/2)$. The constants
$A_{\uparrow,1}$ and $A_{\uparrow,2}$ cannot be derived from the
finite-size corrections for low-lying excitations. Here we only aim
to evaluate the long distance asymptotics of these correlation
functions. Instead of using $N_{u}$ and $N_{b}$ in the oscillation
term, we choose to use $n_{\uparrow}=N_{\uparrow}/L$ and
$n_{\downarrow}=N_{\downarrow}/L$ to elucidate the imbalance in the
densities of spin-up and spin-down fermions. Both sets of variables
are related by the relations $N_{u}=N_{\uparrow}-N_{\downarrow}$ and
$N_{s}=N_{\downarrow}$.

Next we consider the charge density correlation function
$G_{nn}(x,t)$ together with the spin correlation function
$G^{z}(x,t)$. Both of these correlation functions are characterized
by the set of quantum numbers $(\Delta N_{u},\Delta N_{b})=(0,0)$
which allows quantum numbers $\Delta D_{u}\in\mathbb{Z}$ and $\Delta
D_{b}\in\mathbb{Z}$. The leading terms are given by
\begin{eqnarray}
\nonumber G_{nn}(x,t) &=& \langle n(x,t)n(0,0)\rangle \\
&\approx& \nonumber
n^{2}+\frac{A_{nn,1}\cos\left(2\pi(n_{\uparrow}-n_{\downarrow})x\right)}{|x+\mathrm{i}v_{u}t|^{\theta_{1}}}
+\frac{A_{nn,2}\cos\left(2\pi
n_{\downarrow}x\right)}{|x+\mathrm{i}v_{b}t|^{\theta_{2}}}
\\
&&
+\frac{A_{nn,3}\cos\left(2\pi(n_{\uparrow}-2n_{\downarrow})x\right)}{|x+\mathrm{i}v_{u}t|^{\theta_{3}}|x+\mathrm{i}v_{b}t|^{\theta_{4}}},
\\ \nonumber G^{z}(x,t) &=& \langle S^{z}(x,t)S^{z}(0,0)\rangle \\ &\approx& \nonumber (m^{z})^{2}
+\frac{A_{z,1}\cos\left(2\pi(n_{\uparrow}-n_{\downarrow})x\right)}{|x+\mathrm{i}v_{u}t|^{\theta_{1}}}
+\frac{A_{z,2}\cos\left(2\pi
n_{\downarrow}x\right)}{|x+\mathrm{i}v_{b}t|^{\theta_{2}}}
\\
&&
+\frac{A_{z,3}\cos\left(2\pi(n_{\uparrow}-2n_{\downarrow})x\right)}{|x+\mathrm{i}v_{u}t|^{\theta_{3}}|x+\mathrm{i}v_{b}t|^{\theta_{4}}},
\end{eqnarray}
where the operators $n(x,t)$ and $S^{z}(x,t)$ are given in terms of
the fields as
\begin{eqnarray}
n(x,t) &=&
\psi^{\dagger}_{\uparrow}(x,t)\psi_{\uparrow}(x,t)+\psi^{\dagger}_{\downarrow}(x,t)\psi_{\downarrow}(x,t),
\\
S^{z}(x,t) &=&
\frac{1}{2}\left(\psi^{\dagger}_{\uparrow}(x,t)\psi_{\uparrow}(x,t)-\psi^{\dagger}_{\downarrow}(x,t)\psi_{\downarrow}(x,t)\right).
\end{eqnarray}
The critical exponents for asymptotic expressions of $G_{nn}(x,t)$
and $G^{z}(x,t)$ are
\begin{eqnarray}
\theta_{1} &\approx& 2, \qquad  \theta_{2} \approx 2-\frac{2(1-P)}{|\gamma|}, \nonumber  \\
\theta_{3} &\approx& 2+\frac{16P}{|\gamma|}, \qquad  \theta_{4} \approx
2-\frac{2(1-P)}{|\gamma|}+\frac{16P}{|\gamma|}.
\end{eqnarray}
The constant terms for $G_{nn}(x,t)$ and $G^{z}(x,t)$ come from the
choice of quantum numbers $(\Delta D_{u},\Delta D_{b})=(0,0)$. The
second, third and fourth terms arise from the choices $(1,0)$,
$(0,1)$ and $(-1,1)$, respectively.

Finally we consider the pair correlation function $G_{p}(x,t)$. This
correlation function is characterized by the set of quantum numbers
$(\Delta N_{u},\Delta N_{b})=(0,1)$ which allows quantum numbers
$\Delta D_{u}\in\mathbb{Z}+1/2$ and $\Delta D_{b}\in\mathbb{Z}$. The
leading terms are
\begin{eqnarray}
\nonumber G_{p}(x,t) &=&
\langle\psi_{\uparrow}^{\dagger}(x,t)\psi_{\downarrow}^{\dagger}(x,t)\psi_{\uparrow}(0,0)\psi_{\downarrow}(0,0)\rangle
\\ &\approx&
\frac{A_{p,1}\cos\left(\pi(n_{\uparrow}-n_{\downarrow})x\right)}{|x+\mathrm{i}v_{u}t|^{\theta_{1}}|x+\mathrm{i}v_{b}t|^{\theta_{2}}}
+\frac{A_{p,2}\cos\left(\pi(n_{\uparrow}-3n_{\downarrow})x\right)}{|x+\mathrm{i}v_{u}t|^{\theta_{3}}|x+\mathrm{i}v_{b}t|^{\theta_{4}}},
\end{eqnarray}
where the critical exponents are given by
\begin{eqnarray}
\theta_{1} &\approx& \frac{1}{2}, \qquad  \theta_{2} \approx
\frac{1}{2}-\frac{(1-P)}{2|\gamma|}, \nonumber \\ \theta_{3} &\approx&
\frac{1}{2}+\frac{8P}{|\gamma|}, \qquad  \theta_{4} \approx
\frac{5}{2}-\frac{5(1-P)}{2|\gamma|}+\frac{8P}{|\gamma|}.
\end{eqnarray}
The first term in $G_{p}(x,t)$ arises from the choice of quantum
numbers $(\Delta D_{u},\Delta D_{b})=(1/2,0)$, whilst the second
term arises from the choice $(\Delta D_{u},\Delta D_{b})=(1/2,-1)$.

The leading order for the long distance asymptotics of the pair
correlation function $G_{p}(x,t)$  oscillates with wave number
$\Delta k_F$, where $\Delta k_F =\pi(n_{\uparrow}-n_{\downarrow})$.
Meanwhile, the leading order for the spin correlation function
$G^{z}(x,t)$, which can also be thought of as the correlation of the
density difference between spin-up and spin-down fermions,
oscillates twice as fast with wave number $2\Delta k_F$. The
oscillations in $G_{p}(x,t)$ and $G^{z}(x,t)$ are caused by an
imbalance in the densities of spin-up and spin-down fermions, i.e.,
$n_{\uparrow}-n_{\downarrow}$, which gives rise to a mismatch in
Fermi surfaces between both species of fermions. These spatial
oscillations share a similar signature as the Larkin-Ovchinikov (LO)
pairing phase \cite{Larkin1965}. Our findings of the wave numbers
agree with those discovered through DMRG
\cite{Feiguin2007,Tezuka2008,Rizzi2008}, QMC \cite{Batrouni2008} and
mean field theory \cite{Liu2008}. Though from conformal field
theory, we see clearly that the spatial oscillation terms in the
pair and spin correlations are a consequence of Type 3 excitations,
i.e., backscattering for bound pairs and unpaired fermions. A
comparison between our results and the results from numerical
methods in
Refs.~\cite{Feiguin2007,Tezuka2008,Rizzi2008,Batrouni2008} suggest
that the coefficient $A_{p,1}$ is very much larger than the
coefficient $A_{p,2}$ because the frequency of the oscillations in
numerical studies of $G_{p}(x,t)$ is almost identical to
$\pi(n_{\uparrow}-n_{\downarrow})$. This observation also applies to
$G^{z}(x,t)$, where $A_{z,2}$ and $A_{z,3}$ are much smaller when
compared with $A_{z,1}$.


\begin{figure}
\centering
\begin{tabular}{c}
\epsfig{file=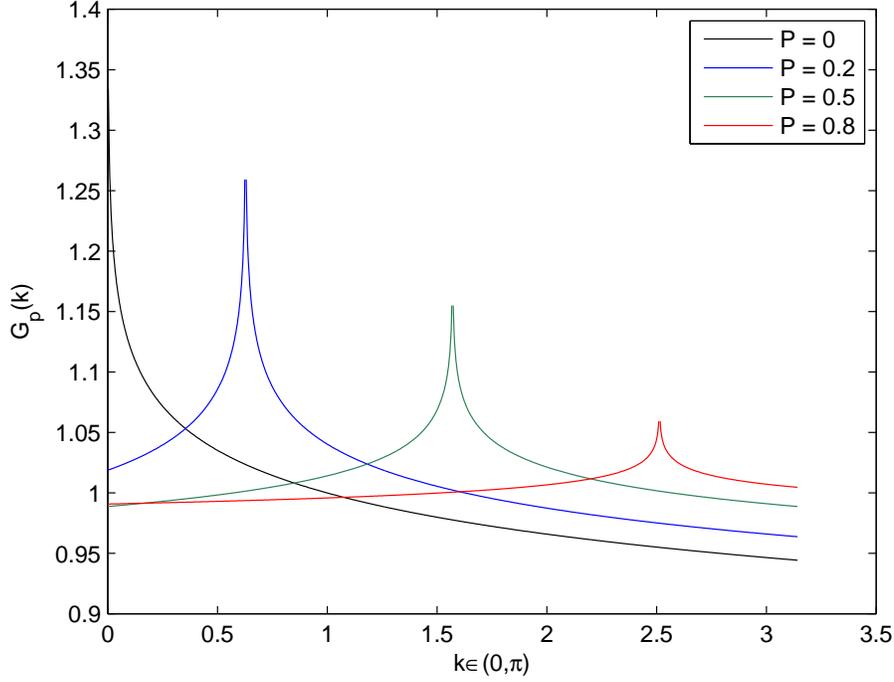,width=0.8\linewidth,clip=}
\end{tabular}\caption{(Color online) This figure shows a plot of the pair correlation function in momentum
space $\widetilde{G}_{p}(k)$ against $k$ for different values of
polarization $P$ when $|\gamma|=10$ and  total linear density $n_{f}=1$. The location of
the peaks are at $k=0$, $0.2\pi$, $0.5\pi$ and $0.8\pi$ when $P=0$,
$0.2$, $0.5$ and $0.8$, respectively.}\label{fig:Gp}
\end{figure}

The correlation functions in momentum space can be derived by taking
the Fourier transform of their counterparts in position space. From
Refs.~\cite{Frahm1991,Hubbardbook}, the Fourier transform of
equal-time correlation functions of the form
\begin{equation}
g(x,t=0^{+}) =
\frac{\exp(ik_{0}x)}{(x-\mathrm{i}0)^{2\Delta^{+}}(x+\mathrm{i}0)^{2\Delta^{-}}},
\end{equation}
where $\Delta^{\pm}=\Delta_{u}^{\pm}+\Delta_{b}^{\pm}$ is given by
\begin{equation}
\widetilde{g}(k\approx
k_{0})\sim[\mathrm{sign}(k-k_{0})]^{2s}|k-k_{0}|^{\nu}.
\end{equation}
The conformal spin of the operator is $s=\Delta^{+}-\Delta^{-}$ and
the exponent $\nu$ is expressed in terms of the conformal dimensions
as $\nu=2(\Delta^{+}+\Delta^{-})-1$.

Hence the equal time correlation functions near the singularities
$k_{0}$ for the one particle Green's function, charge density, spin
and bound pairs are
\begin{eqnarray}
\widetilde{G}_{\uparrow}(k) &\sim&
[\mathrm{sign}(k-\pi(n_{\uparrow}-2n_{\downarrow}))]^{2s_{\uparrow}}|k-\pi(n_{\uparrow}-2n_{\downarrow})|^{\nu_{\uparrow}},
\label{eq:corr_mom1}
\\ \widetilde{G}_{nn}(k) &\sim& [\mathrm{sign}(k-2\pi(n_{\uparrow}-n_{\downarrow}))]^{2s_{nn}}|k-2\pi(n_{\uparrow}-n_{\downarrow})|^{\nu_{nn}},
\\ \widetilde{G}^{z}(k) &\sim& [\mathrm{sign}(k-2\pi(n_{\uparrow}-n_{\downarrow}))]^{2s_{z}}|k-2\pi(n_{\uparrow}-n_{\downarrow})|^{\nu_{z}},
\\ \widetilde{G}_{p}(k) &\sim&
[\mathrm{sign}(k-\pi(n_{\uparrow}-n_{\downarrow}))]^{2s_{p}}|k-\pi(n_{\uparrow}-n_{\downarrow})|^{\nu_{p}},
\label{eq:corr_mom4}
\end{eqnarray}
where the exponents are given by
\begin{eqnarray}
&& 2s_{\uparrow} \approx
1+\frac{4P}{|\gamma|}-\frac{(1-P)}{|\gamma|},\qquad \nu_{\uparrow}
\approx \frac{1}{2}+\frac{8P}{|\gamma|}-\frac{(1-P)}{2|\gamma|},
\\ && 2s_{nn} = 2s_{z} \approx 0,\qquad \nu_{nn} = \nu_{z}
\approx 1, \\ && 2s_{p} \approx 0,\qquad \nu_{p} \approx
-\frac{(1-P)}{2|\gamma|}.
\end{eqnarray}
We would like to stress that the momentum space correlation
functions derived in Eqs.~\eqref{eq:corr_mom1}--\eqref{eq:corr_mom4}
are only accurate when the momenta $k$ are within the proximity of
the wave numbers $k_{0}$, i.e., when $k\approx k_{0}$.
FIG.~\ref{fig:Gp} plots $\widetilde{G}_{p}(k)$ against $k$ as
polarization $P$ varies between 0 to 0.8. This figure is in
qualitative agreement with the ones given in
Refs.~\cite{Feiguin2007,Batrouni2008,Rizzi2008}. We stress again
that our plot is accurate only within the vicinity of the
singularity, i.e., when $k$ approaches
$\pi(n_{\uparrow}-n_{\downarrow})$. We plotted
$\widetilde{G}_{p}(k)$ for the entire domain $k\in(0,\pi)$ so that
readers can visualize the curves more easily.

\section{Conclusion}\label{sec:Conclusion}
In conclusion, we investigated various zero-temperature correlation
functions for the spin-1/2 Fermi gas  with attractive
interaction. We derived the finite-size corrections for ground state
and low-lying excitations of the model. Using conformal field
theory, critical exponents of the correlation functions were given
in terms of polarization and interaction strength. We found that the
leading terms of the pair correlation function and the spin
correlation function oscillate with frequencies
$\pi(n_{\uparrow}-n_{\downarrow})$ and
$2\pi(n_{\uparrow}-n_{\downarrow})$, respectively. We also found
that backscattering between the Fermi points of bound pairs and
unpaired fermions results in a 1D analog of the FFLO state and
displays a microscopic origin of the FFLO nature. Furthermore, we
showed that there is a peak in the pair correlation function in
momentum space at $k=\pi(n_{\uparrow}-n_{\downarrow})$ which
confirms the oscillation frequency.

In the spin polarized phase, these correlation functions exhibit
spatial oscillations with a power-law decay. This critical behaviour
can be viewed as an analogy to long range order in 1D, i.e., the
power law decay of the pair correlation function which is regarded
as evidence of a superconducting/superfluid state. We also like to
mention that from the dressed charge formalism, the asymptotic
behavior of the correlation functions derived in this paper can be
numerically obtained with high accuracy for arbitrary interaction
strength. Additionally, by considering weakly perturbed inter-tube
interactions or inter-lattice interactions (1D fermionic Hubbard
model), quasi-1D correlations in the spin polarized phase can be
calculated from perturbation theory \cite{Bogoliubov1989}. This
provides a promising opportunity to estimate the critical
temperature for high-Tc superconductors/superfluids by studying 1D
to 3D trapped cold atoms.

\begin{acknowledgments}
This work is supported by the Australian Research Council.  We 
thank M. T. Batchelor, F. H. L. Essler, F.  Heidrich-Meisner and K. Sakai for helpful
discussions.  XWG  thanks T.-L. Ho for stimulating discussions to initiate this topic  and acknowledges the Ohio State University for their
kind hospitality.
\end{acknowledgments}



\begin{thebibliography}{99}

\bibitem{Fulde1964} P. Fulde and R. A. Ferrell, Phys. Rev. {\bf
135}, A550 (1964)

\bibitem{Larkin1965} A. I. Larkin and Yu. N. Ovchinnikov, Sov. Phys.
JETP {\bf 20}, 762 (1965)

\bibitem{Yang1967} C. N. Yang, Phys. Rev. Lett. {\bf 19}, 1312 (1967)

\bibitem{Gaudin1967} M. Gaudin, Phys. Lett. A {\bf 24}, 55 (1967)

\bibitem{Orso} G. Orso, Phys. Rev. Lett. {\bf 98}, 070402 (2007)

\bibitem{HuHui} H. Hu, X.-J. Liu and P. D. Drummond, Phys. Rev. Lett. \textbf{98}, 070403 (2007)

\bibitem{Feiguin2007} A. E. Feiguin and F. Heidrich-Meisner, Phys.
Rev. B {\bf 76}, 220508(R) (2007)

\bibitem{Tezuka2008} M. Tezuka and M. Ueda, Phys. Rev. Lett. {\bf
100}, 110403 (2008)

\bibitem{Rizzi2008} M. Rizzi, M. Polini, M. A. Cazalilla, M. R.
Bakhtiari, M. P. Tosi and R. Fazio, Phys. Rev. B {\bf 77}, 245105
(2008)

\bibitem{Meisner} F. Heidrich-Meisner, A. E. Feiguin, U. Schollw\"{o}ck and W. Zwerger, Phys. Rev. A {\bf 81}, 023629 (2010)

\bibitem{Luscher} A. L\"{u}scher, R. M. Noack and A. M. L\"{a}uchli, Phys. Rev. A {\bf  78}, 013637 (2008)

\bibitem{Batrouni2008} G. G. Batrouni, M. H. Huntley, V. G. Rousseau
and R. T. Scalettar, Phys. Rev. Lett. {\bf 100}, 116405 (2008)

\bibitem{Kinnunen2006} J. Kinunnen, L. M. Jensen and P.
T\"{o}rm\"{a}, Phys. Rev. Lett. {\bf 96}, 110403 (2006)

\bibitem{Liu2008} X.-J. Liu, H. Hu and P. D. Drummond, Phys. Rev. A
{\bf 78}, 023601 (2008)

\bibitem{Cooper}J. M. Edge and N. R. Cooper, Phys. Rev. Lett. {\bf 103}, 065301 (2009)

\bibitem{Zhao2008} E. Zhao and W. V. Liu, Phys. Rev. A {\bf 78},
063605 (2008)

\bibitem{Yang2001} K. Yang, Phys. Rev. B {\bf 63}, 140511(R) (2001)

\bibitem{Bogoliubov1988} N. M. Bogoliubov and V. E. Korepin, Mod.
Phys. Lett. B {\bf 1}, 349 (1988)

\bibitem{Bogoliubov1989} N. M. Bogoliubov and V. E. Korepin, Int.
J. Mod. Phys. B {\bf 3}, 427 (1989)

\bibitem{Bogolyubov1990} N. M. Bogolyubov and V. E. Korepin, Theor.
Math. Phys. {\bf 82}, 231 (1990)

\bibitem{Bogolyubov1992} N. M. Bogolyubov and V. E. Korepin,
Proceedings of the Steklov Institute of Mathematics {\bf 2}, 47
(1992)

\bibitem{Guan2007} X. W. Guan, M. T. Batchelor, C. Lee and M. Bortz,
Phys. Rev. B {\bf 76}, 085120 (2007)

\bibitem{Mueller}M. Casula, D M. Ceperley and E. J. Mueller, Phys. Rev. A {\bf 78}, 033607 (2008)

\bibitem{Kakashvili} P. Kakashvili and C. J. Bolech, Phys. Rev. A, \textbf{79}, 041603(R) (2009)

\bibitem{Wadati} T. Iida and M. Wadati, J. Phys. Soc. Jpn, {\bf 77},  024006 (2008)

\bibitem{Penc} F. Woynarovich and K. Penc, Z. Phys. B {\bf 85}, 269 (1991)

\bibitem{Erhai} E. Zhao, X. W. Guan, W. V. Liu, M. T. Batchelor and M.
Oshikawa, Phys. Rev. Lett. \textbf{103}, 140404 (2009).

\bibitem{Liao2009} Y. Liao, A. S. C. Rittner, T. Paprotta, W. Li, G. B. Partridge,
R. G. Hulet, S. K. Baur and E. J. Mueller, Nature {\bf 467}, 567
(2010)

\bibitem{Y-Y} C. N. Yang and C. P. Yang, J. Math. Phys. {\bf 10}, 1115 (1969)

\bibitem{Takahashi} M. Takahashi, {\it Thermodynamics of
  One-Dimensional Solvable Models}, Cambridge University Press (1999)

\bibitem{Lai1971} C. K. Lai, Phys. Rev. Lett.
{\bf 26}, 1472 (1971)

\bibitem{Lai1973} C. K. Lai, Phys. Rev. A {\bf 8}, 2567 (1973)

\bibitem{Schlottmann1993} P. Schlottmann, J. Phys.: Condens. Matter
{\bf 5} 5869 (1993)


\bibitem{Belavin1984} A. A. Belavin, A. M. Polyakov and A. B.
Zamolodchikov, Nucl. Phys. B {\bf 241}, 333 (1984)

\bibitem{Blote1986} H. W. Bl\"{o}te, J. L. Cardy and M. P.
Nightingale, Phys. Rev. Lett. {\bf 56}, 742 (1986)

\bibitem{Affleck1986} I. Affleck, Phys. Rev. Lett. {\bf 56}, 746 (1986)



\bibitem{Guan2010} X. W. Guan, J.-Y. Lee, M. T. Batchelor, X.-G. Yin
and S. Chen, Phys. Rev. A {\bf 82}, 021606(R) (2010)

\bibitem{Cardy1986} J. L. Cardy, Nucl. Phys. B {\bf 270} [FS16], 186 (1986)

\bibitem{Izergin1989} A. G. Izergin, V. E. Korepin and N. Yu
Reshetikhin, J. Phys. A: Math. Gen. {\bf 22}, 2615 (1989)

\bibitem{Woynarovich1989} F. Woynarovich, J. Phys. A {\bf 22}, 4243
(1989)

\bibitem{Kawakami1991} N. Kawakami and S. K. Yang, J. Phys. C {\bf
3}, 5983 (1991)

\bibitem{Frahm1991} H. Frahm and V. E. Korepin, Phys. Rev. B {\bf
43}, 5653 (1991)

\bibitem{Hubbardbook} F. H. L. Essler, H. Frahm, F. G\"{o}hmann, A.
Kl\"{u}mper and V. E. Korepin, {\it The One-Dimensional Hubbard
Model}, Cambridge University Press (2005)

\end{thebibliography}
\end{document}